\documentclass{comjnl}

\usepackage{graphicx}
\usepackage{caption}
\usepackage{subcaption}
\usepackage{siunitx}
\usepackage{amsmath}
\usepackage[sorting=none,style=numeric-comp]{biblatex}
\bibliography{myArticles}

\begin{document}

\title[Light Story]{Manipulating organic semiconductor morphology with visible light}
\author{Michael Korning S\o rensen$^1$, Anders Skovbo Gertsen$^1$, Rocco Peter Fornari$^1$, Binbin Zhou$^2$, Peter Uhd Jepsen$^2$, Edoardo Stanzani$^1$, Shinhee Yun$^1$, Marcial Fernández Castro$^1$,  Matthias Schwartzkopf$^3$, Alexandros Koutsioubas$^4$, Piotr de Silva$^1$, Moises Espindola Rodriguez$^1$, Luise Theil Kuhn$^1$, and Jens Wenzel Andreasen*}

\affiliation{Department of Energy Conversion and Storage, Technical University of Denmark, Fysikvej 310, Kgs. Lyngby, 2800, Denmark}
\affiliation{Department of Photonics Engineering, Technical University of Denmark, \O rsted Plads, Kgs. Lyngby, 2800, Denmark}
\affiliation{Deutsches Elektronen-Synchrotron (DESY), Notkestraße 85, D-22607 Hamburg, Germany}
\affiliation{Jülich Centre for Neutron Science (JCNS) at Heinz Maier-Leibnitz Zentrum (MLZ), Forschungszentrum Jülich GmbH, Lichtenbergstr. 1, 85748 Garching, Germany}

 \email{jewa@dtu.dk}

\shortauthors{M.K. S\o rensen}

\keywords{Organic semiconductor, P3HT, poly(3-hexylthiophene), roll-to-roll, slot-die coating, LED light treatment, GIWAXS, GISAXS, GISANS, SAXS, WAXS, AFM, Absorbance, DFT, Solvent evaporation, coarse-grained, MARTINI, molecular dynamics, THz Spectroscopy, Organic solar cells }

\begin{abstract}

We present a method to manipulate the final morphology of roll-to-roll slot-die coated poly(3-hexylthiophene) (P3HT) by optically exciting the p-type polymer in solution while coating. Our results provide a comprehensive picture of the entire knowledge chain, from demonstrating how to apply our method to a fundamental understanding of the changes in morphology and physical properties induced by exciting P3HT while coating. By combining results from density functional theory and molecular dynamics simulations with a variety of X-ray experiments, absorption spectroscopy, and THz spectroscopy, we demonstrate the relationship between morphology and physical properties of the thin film. Specifically, in P3HT films excited with light during deposition, we observe changes in crystallinity and texture with more face-on orientation and increased out-of-plane charge mobility.

\end{abstract}

\maketitle
State-of-the-art organic photovoltaics (OPV)  achieve efficiencies above 18 $\%$ \cite{Liu2018}, which is competitive with their silicon counterparts \cite{Green2021}. However, such record solar cell efficiencies are obtained from small areas, less than 1 cm$^2$, and fabricated with deposition methods, e.g., spin-coating, that are not compatible with large-scale production. In general, the best performing OPVs fabricated with large-scale methods such as roll-to-roll coating are trailing behind with approximately half the efficiency, i.e., 5 - 10 $\%$ \cite{GertCastro2020,Wang2020}. Developing methods to close this gap is essential for a greener future.

\begin{figure*}[t]
\centering
  \includegraphics[height=12.8cm]{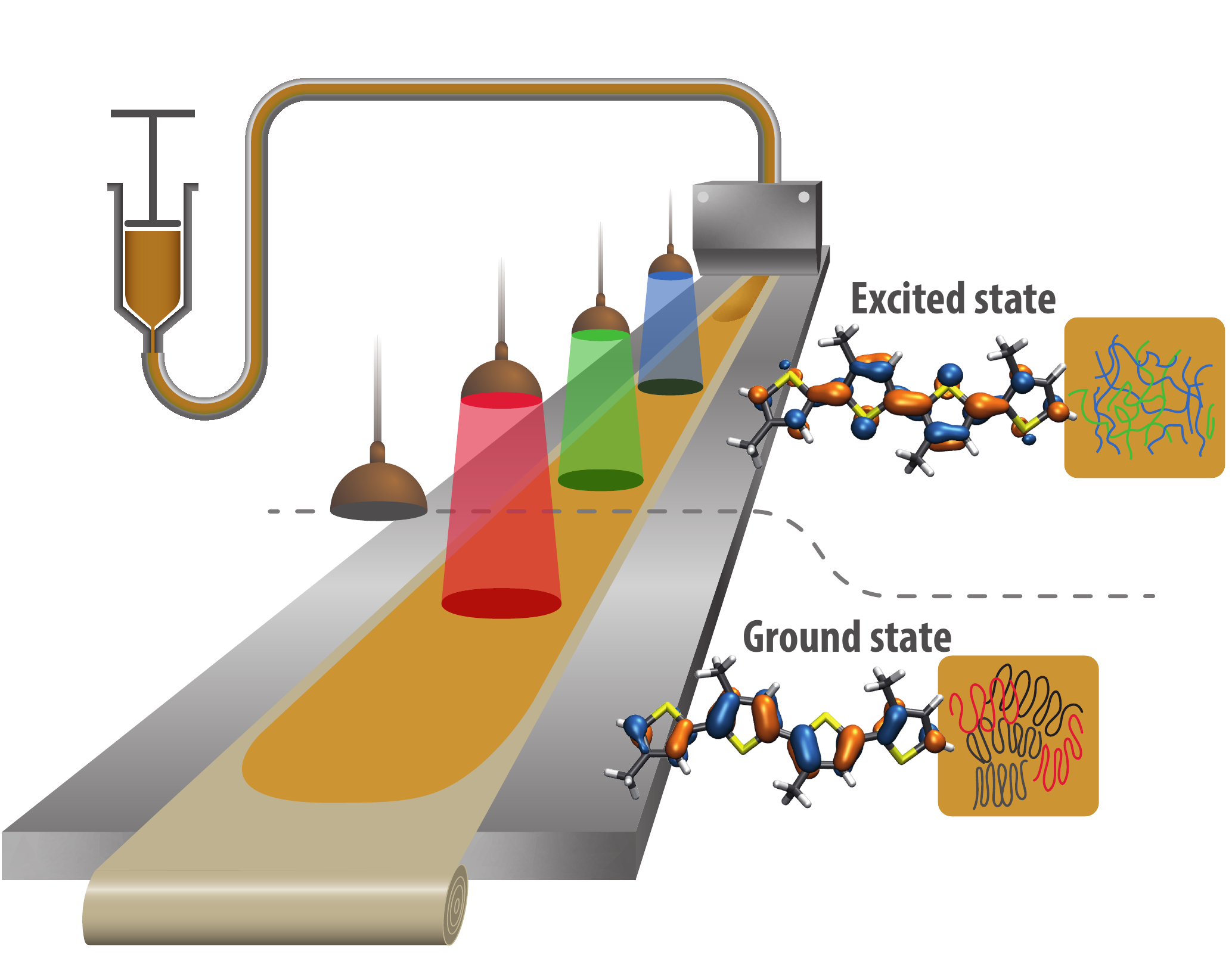}
  \caption{Illustration of the fabrication method where P3HT in solution is slot-die coated on a moving substrate either in the dark (no light), or illuminated with red light, green light, or blue light. The HOMO and LUMO molecular orbitals, obtained from DFT calculations of a poly(3-methylthiophene) (P3MT) 4-mer, represent the dominant bonding pattern in the ground and excited state. When exposed to no light or red light, P3HT is expected to remain in the ground state, as opposed to green light and blue light which possess enough energy to excite the polymer. The presence or absence of excited polymers during deposition will influence the final morphology of the thin film as conceptually depicted in the yellow squares.  }
  \label{fgr:rRGB}
\end{figure*}

The conjugated polymer poly(3-hexylthiophene) (P3HT) is often used as  electron donor in  donor-acceptor heterojunctions for OPVs \cite{P3HTnfa2019,holiday2016,Baran2017, krebs2014}. It is generally understood that the morphology of thin films of P3HT plays an important role in determining the physical and electric properties of the film \cite{ludwigs2014}. Controlling the final morphology of P3HT films depends on several process parameters which include coating method, temperature, post-annealing treatments, solvents, additives, concentration, choice of substrate, gas flow above the drying film, laser treatment of film after coating, molecular weight, and regioregularity \cite{rpm2005,tobias2017b,p3ht_temp3,krebs2014anneald, roth2019,Kajiya2018,Feng2017Laser-inducedOf,P3HT_MW,medom1999,RRp3ht}. Thus, optimizing the fabrication method is paramount to achieve a specific morphology and  desired physical properties for best thin film performance. 

This paper presents a method to manipulate the  morphology of P3HT thin films by optically exciting the polymer in solution with visible light from light-emitting diodes (LEDs) while roll-to-roll slot-die coating. Four different treatments were chosen to match energies both below and above the excitation energy levels of P3HT in solution and in thin film \cite{Rahimi2014}. Specifically, either no illumination (i.e., dark), or LEDs emitting red light (625 nm), green light (525 nm), or blue light (465 nm)  illuminated the films while being coated as illustrated in Figure \ref{fgr:rRGB}. It is known from density functional theory (DFT) calculations \cite{Roseli2017} that, when exciting P3HT with visible light, i.e. going from the ground state to the excited state, the electron density changes to form double bonds between neighbouring thiophene rings as depicted in Figure \ref{fgr:rRGB}. As discussed in the next section, this change in the bonding pattern planarizes the polymer backbone. As a consequence, the final morphology is affected when exciting the polymer during coating, which is the subject of study in this paper. To probe the morphological changes caused by the four treatments, we use Grazing-Incidence Wide-Angle X-ray Scattering (GIWAXS) on the coated films. To elucidate the physical origin of the changes, DFT and solvent evaporation molecular dynamics (MD) are used to simulate the effect of excitation on single polymer chains and thin films. The full picture of morphological changes is followed by a thorough investigation of change of physical properties in the thin film. Here, the  UV-VIS absorption spectra are analyzed, followed by THz spectroscopy measuring the thin-film photo-conductivity. Finally, the consequence of light treating a complete OPV device (P3HT:O-IDTBR) is  discussed.

\section*{Excitation of P3HT in solution}
\begin{figure}[t]
\centering
 \includegraphics [width=\columnwidth] {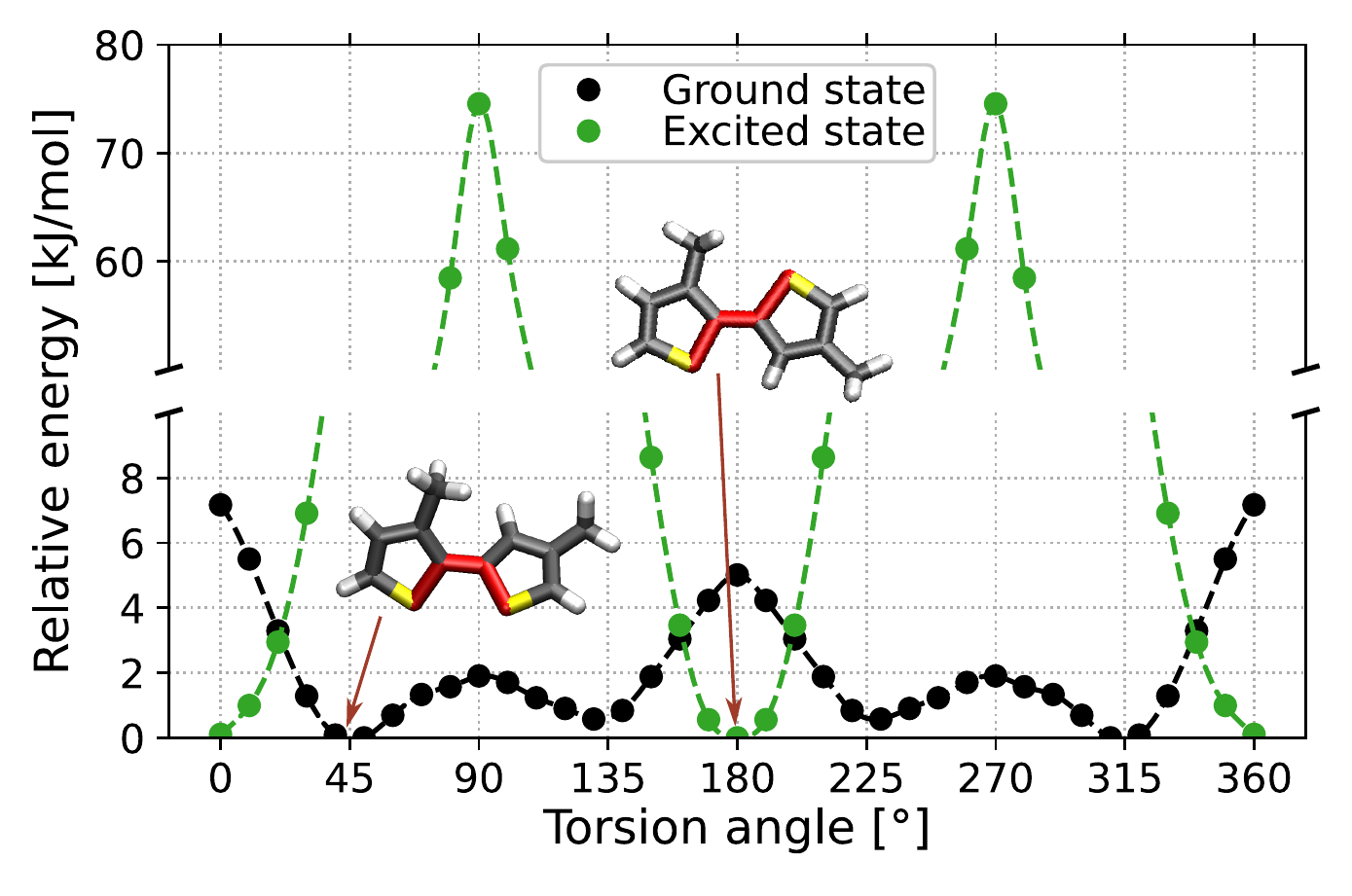}
 \caption{Torsional profiles of the ground state and  first excited state of a 3MT (3-methylthiophene) dimer. Each data point is obtained by constraining the dihedral angle between monomers to a given value and optimizing the remaining degrees of freedom. The energy is relative to the minimum of each energy profile. Geometries corresponding to the ground and excited state minima are shown in the insets.}
 \label{fig:torsion}
\end{figure}

The main structural difference between the ground state (GS) $S_0$ and the first excited state (ES) $S_1$ of P3HT lies in the dihedral angles between neighboring thiophene rings. It is well known that the GS of thiophene oligomers preferentially adopts twisted geometries with equilibrium dihedral angles around 45$^{\circ}$, whereas the excited states are perfectly planar \cite{Roseli2017}.
The torsional profiles obtained from relaxed scans performed on 3-methylthiophene (3MT) dimers are shown in Figure \ref{fig:torsion}. In the GS, the two minima around 45$^{\circ}$ and 135$^{\circ}$ are separated by relatively low barriers of ~2 kJ/mol at 90$^{\circ}$ and ~5 kJ/mol at 180$^{\circ}$, which can be overcome at room temperature ($RT \sim$ 2.5 kJ/mol at 300K). This ease of rotation between monomers confers a rather large conformational freedom to P3HT chains in the GS. In contrast, the ES torsional profile has minima at 0$^{\circ}$ (\textit{cis} conformation) and 180$^{\circ}$ (\textit{trans} conformation) separated by much higher barriers of ~75 kJ/mol, which essentially lock the dimer in one of the two co-planar conformations. This difference between the GS and ES torsional profiles is due to the electronic structure. As in many conjugated polymers, the $S_0$ to $S_1$ transition is due mainly to HOMO to LUMO excitation, as shown in Figure \ref{fgr:rRGB}, and is accompanied by an aromatic to quinoid electronic rearrangement \cite{Roseli2017}. The electronic density shifts from the intra-ring aromatic system to the inter-ring bonds, which acquire a more pronounced double bond character that inhibits rotation. It is reasonable to assume that the gas-phase torsional potential of the excited 3MT dimer, presented in Figure \ref{fig:torsion}, is a good approximation of P3HT in solution. From transmission Small Angle X-ray Scattering (SAXS) it is observed that keeping P3HT in the dark, i.e. in the ground state, the sizes of P3HT aggregates slowly increase over time. Conversely, when excited by green light, i.e. with several molecules in the excited state, the size of aggregates in solution remains constant as experimentally shown in Supplementary, Figures 2 and 3.

\section*{Excitation of P3HT during coating}

\begin{figure*}[ht]
\centering
  \includegraphics[height=10cm]{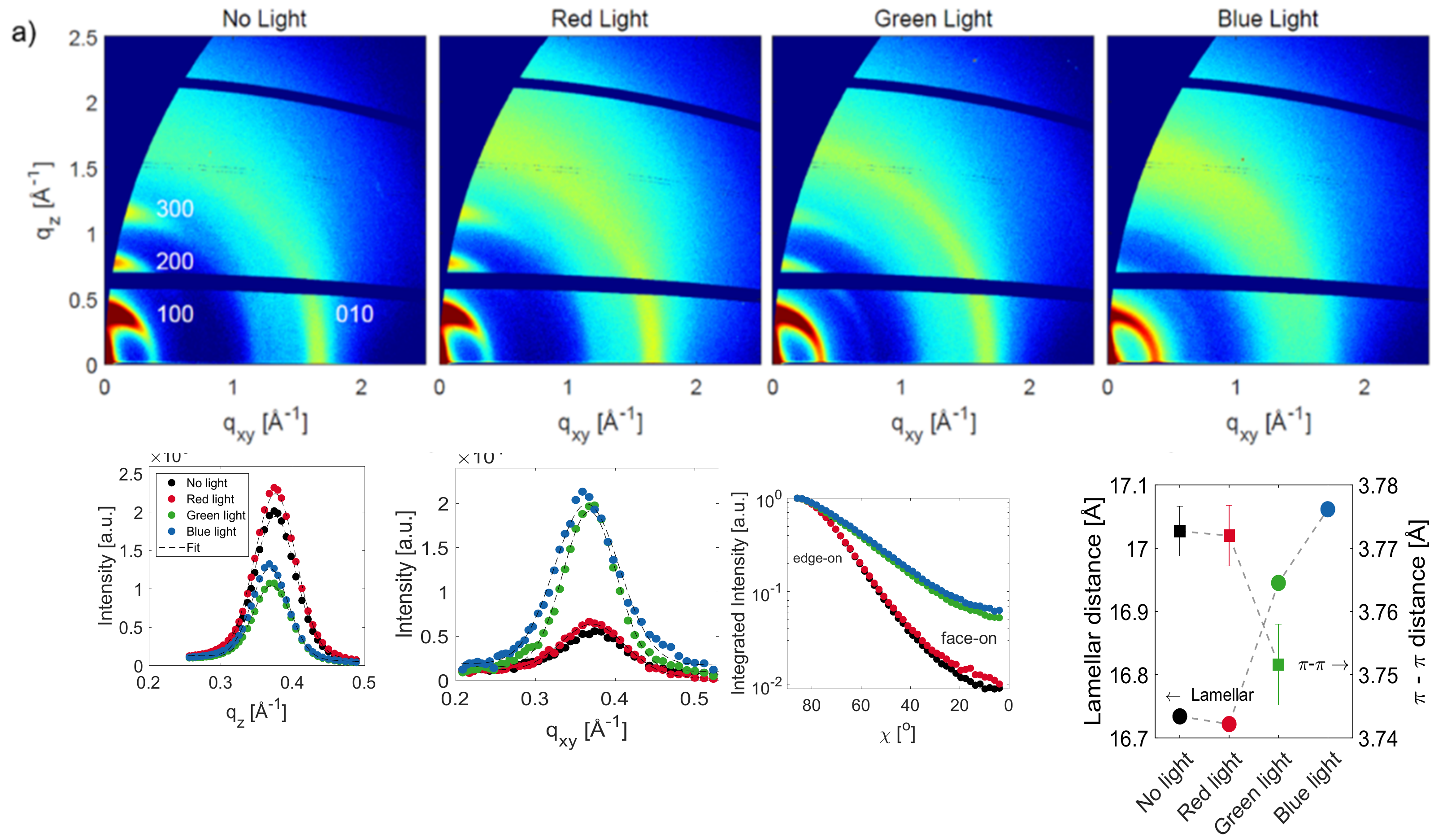}
  \caption{ a) GIWAXS data with Ewald sphere correction of P3HT slot-die coated on top of silicon with varying light treatments. Lamellar stack (100) and $\pi - \pi$ stack (010) signals are indicated with white text in the left panel. The colour scale is logarithmic and equal for the four data sets shown. b) Out-of-plane (q$_z$) line integration of lamellar peak and Gaussian fit. c) In-plane (q$_{xy}$) line integration of lamellar peak and Gaussian fit. d) Integrated lamellar signal as a function of $\chi$, where 0$^{\circ}$ corresponds to in-plane and 90$^{\circ}$ corresponds to out-of-plane orientation of the lamellar stacking direction, indicating face-on and edge-on orientation of the polymer planes with respect to the substrate. e) Out-of-plane lamellar and in-plane $\pi-\pi$ packing distances for the four treatments. Lamellar distances are indicated with circular symbols (error bars are smaller than the symbols) and refer to the left axis, whereas the $\pi-\pi$ distances are indicated with squares and refer to the right axis. As a consequence of not having any in-plane $\pi-\pi$ stacking peak in the blue light sample (see Supplementary, Figure 5), no value is shown here. } 
  \label{fgr:2D}
\end{figure*}

The 2D GIWAXS scattering data obtained from the four samples are shown in Figure \ref{fgr:2D}a,  where the lamellar stack (100, 200, and 300) and $\pi - \pi$ stack (010) signals are indicated. Qualitatively, we observe that the P3HT chains are predominantly oriented edge-on with respect to the substrate with a degree of orientational distribution. To quantitatively evaluate the morphological changes, the lamellar stack (100) and the $\pi - \pi$ stack (010) peaks are fitted by Gaussian distributions along the surface normal (q$_z$) and in the substrate plane (q$_{xy}$) with suitable background corrections $(q = 4\pi\sin(\theta)/\lambda$, where $\theta$ is half the scattering angle, and $\lambda$ is the X-ray wavelength).

In Figure \ref{fgr:2D}b, the intensities of the lamellar (100) reflections are shown as a function of the reciprocal space out-of-plane distance $q_z$ (i.e., along the normal to the substrate) for different light treatments. The observed intensity is approximately halved in the films treated with green and blue light compared to those with no and red light. Additionally, the center of the peak is shifted to lower $q_z$ for the two lower intensity profiles. In real space, these observations are interpreted as an increase of the lamellar packing distance, a decrease of overall crystallinity, and a decrease of edge-on oriented aggregates when treated with green and blue light compared to the no and red light treatment.

In Figure \ref{fgr:2D}c, the intensities of the lamellar (100) reflections are shown as a function of the reciprocal space in-plane coordinate $q_{xy}$ (i.e. in the substrate plane) for different light treatments. The observed intensities increase drastically in the films treated with green and blue light, approximately four times more intense than the other two treatments. This observation indicates a significant increase of the fraction of aggregates oriented face-on with respect to the substrate.

\begin{figure*}[t]
\centering
\includegraphics[width=\textwidth]{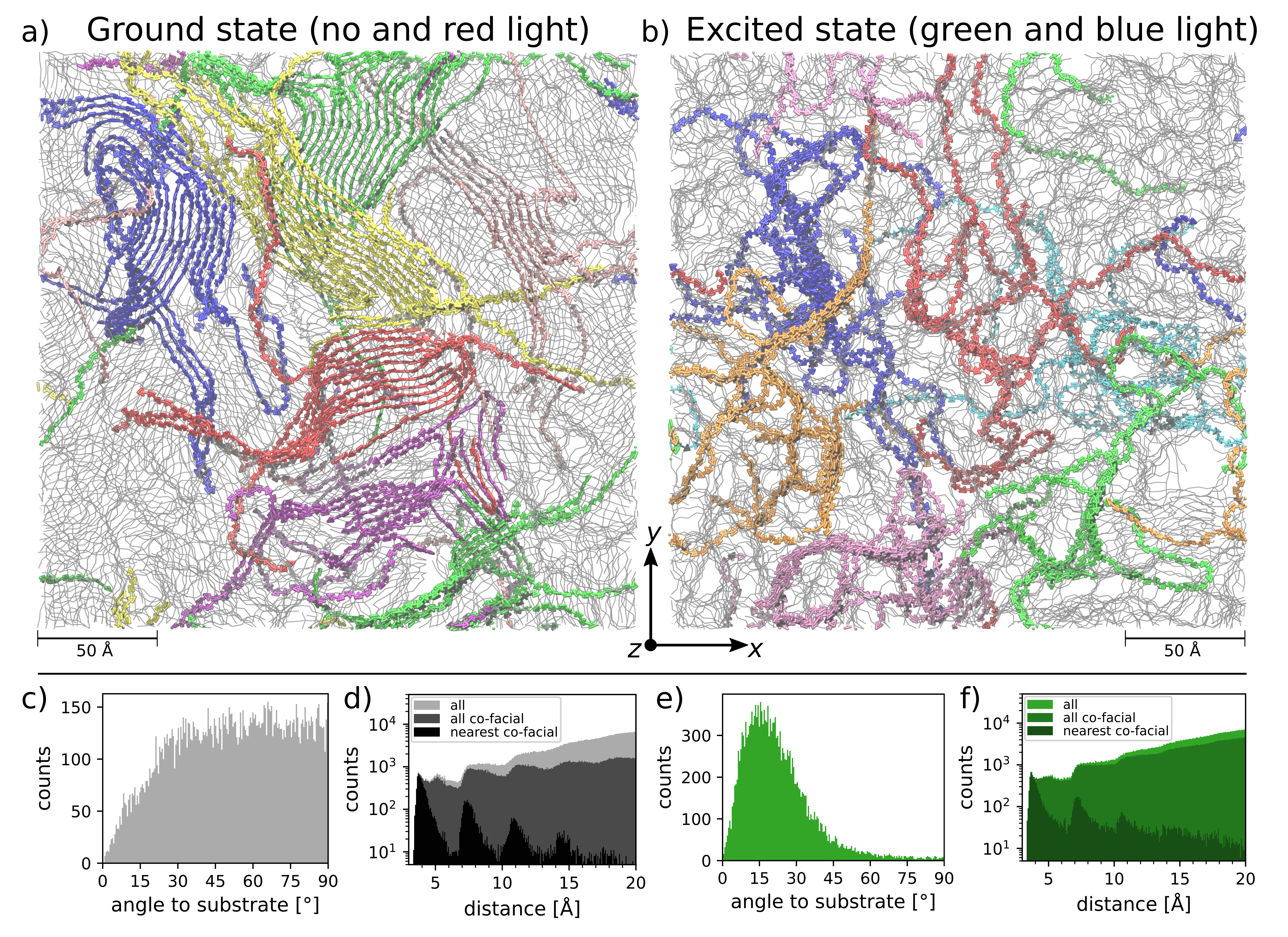}
\caption{Thin film morphologies obtained from coarse-grained MD simulations of solvent evaporation using the GS (a) and ES (b) force field. The coloured aggregates are manually selected groups of adjacent chains with significant $\pi-\pi$ interaction. c), e) Distributions of angles between thiophene ring planes and \textit{xy} plane in the GS and ES films. d), f) Distributions of distances between ring centers in the GS and ES films. Couples of rings are considered co-facial if the angle between their planes is smaller than 30$^{\circ}$. The darker distributions include only the nearest co-facial neighbor of each ring.}
\label{fig:MDmorph}
\end{figure*}

To quantify the distribution of orientations of the lamellar with respect to the substrate, 45$^{\circ}$ azimuthal integrations across the lamellar peak were performed. each with an arc-width of 2$^{\circ}$, going from $I(q)_{\chi_0} \parallel \hat{q}_z $ to  $ I(q)_{\chi_{90}} \bot  \hat{q}_z$. A Gaussian distribution is fitted to the intensity as function of $\chi$ for each integration. The results are shown in Figure \ref{fgr:2D}d (all fits are shown in Supplementary, Figures 6-9). We observe that the distributions of $\chi$ of the no light and red light films are very similar to each other and strongly differ from those of the films treated with green and blue light, which are also similar to each other. This yields a predominant edge-on orientation of the crystallites. However, the samples treated with green and blue light possess a significantly higher fraction of face-on polymer chains.

In Figure \ref{fgr:2D}a it can be seen that the orientation of the $\pi-\pi$ (010) stacking signal changes from being predominantly in-plane, $q_{xy}$, towards a higher degree of out-of-plane, $q_z$, as the treatment increases in energy from no light to blue light. The in-and out-of-plane line integrations of the $\pi-\pi$ stacking signals are shown in Supplementary, Figure 5. In Figure \ref{fgr:2D}e the real space in-plane $\pi-\pi$ stack (010) and the out-of-plane lamellar stack (100) are shown. The in-plane lamellar stacking distances become larger when exposed to green and blue light, indicating less efficient packing.

Next, we report the results of DFT calculations and MD simulations and discuss the effect of excitation on the morphology of P3HT thin films. Photoexcitation of P3HT aggregates first produces a "hot" exciton, delocalized over few monomers, that quickly (on a time scale of 1-10 ps) decays to more delocalized "cold" exciton that has longer lifetimes in the range of 0.1-1.0 ns \cite{p3ht_temp}. This transition to the "cold" exciton is accompanied by torsional relaxation into a more planar geometry that is a minimum of the excited state potential (see Figure \ref{fig:torsion}). It is therefore reasonable to assume that continuous photoexcitation with green or blue light has the effect of keeping significant portions of the polymer chains planar on nanosecond timescales, locking the dihedral angles in either the \textit{cis} or \textit{trans} conformation and preventing rotation between monomers.
In the MD simulation of solvent evaporation with light, for simplicity, we have chosen to keep the molecules in the ES force field all the time. This is, of course, a drastic approximation, but such an extreme model is nonetheless useful to understand the effect of excitation on morphology.
The thin film morphologies obtained from the coarse-grained MD simulations of solvent evaporation without and with light (using the GS and ES force field, respectively) are shown in Figure \ref{fig:MDmorph}. The two morphologies are strikingly different: the GS sample consists mainly of lamellar aggregates typical of P3HT, consisting of roughly 10 $\pi$-stacked chain sections of about 25 monomers. The ES film appears much more disordered from the top, i.e., when viewed along the \textit{z} direction normal to the substrate (right panel of \ref{fig:MDmorph}). A more detailed examination reveals that in the ES sample, most molecules form $\pi-\pi$ stacks in the \textit{z} direction but do not form large lamellae as in the GS sample. Instead, each chain tends to participate in multiple smaller aggregates with short sections of 5-10 monomers. The GS sample presents more efficient packing than the ES sample, with final densities of 1.13 and 1.05 g/mL, respectively.

\begin{figure*}[tb]
\centering
  \includegraphics[width=1\textwidth]{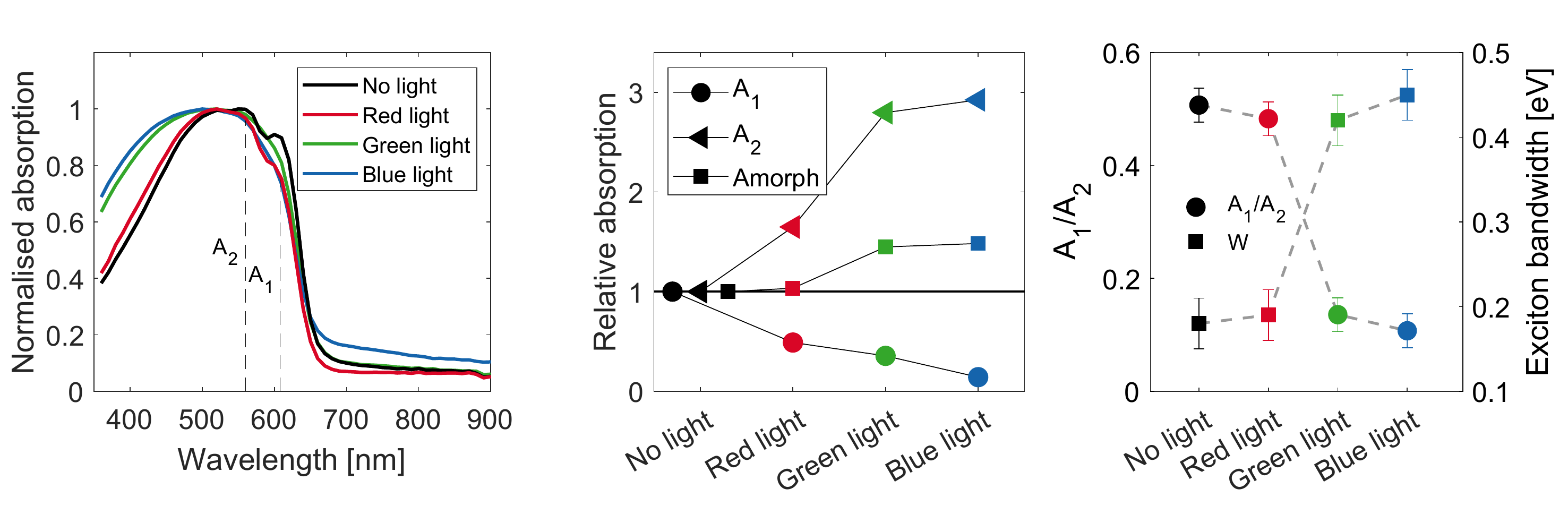} 
  \caption{a) Normalized UV-VIS absorption spectra of P3HT films obtained with four different light treatments. Dashed black lines indicate the locations of the absorption edges: $A_{1}$, $A_{2}$. b) Relative integrated absorption in relation the no light samples for each edge as a function of light treatment. Both $A_{1}$, $A_{2}$ are modeled with two a Gaussian distribution. c) ${A_{1}}/{A_{2}}$ ratio and exciton bandwidth $W$ calculated from eq. (\ref{eq:AA}).}
\label{fig:Absorbe}
\end{figure*}

The distributions of angles between the thiophene rings and the substrate (\textit{xy} plane) are shown in Figure \ref{fig:MDmorph}c and \ref{fig:MDmorph}e. In the GS film, the distribution of angles is rather flat, indicating almost isotropic orientation, with  21$\%$ of the rings face-on (angle $\leq $30$^{\circ}$). In contrast, in the ES film 77\% of the rings are within 30$^{\circ}$ of the \textit{xy} plane: this strong prevalence of face-on orientation, although possibly exaggerated by keeping all chains excited all the time, the trend is consistent with the GIWAXS data in Figure \ref{fgr:2D}.
Figure \ref{fig:MDmorph}d and \ref{fig:MDmorph}f are the distributions of distances between all ring centers and two subsets thereof (see legend): those which are co-facial to each other (angle $\leq$ 30$^{\circ}$) and only the nearest co-facial neighbors. The first peak at around 3.75 Å is composed exclusively of $\pi-\pi$ stacked nearest neighbors and is very similar in the two samples. Similar $\pi-\pi$ stacking distances are observed with GIWAXS as shown in Figure \ref{fgr:2D}e, although we cannot detect its average decrease in the MD morphologies. The successive peaks are at multiples of this distance and are less pronounced in the ES sample. The GS sample presents pronounced co-facial non-neighboring peaks at 7.5, 11.25, 15.0, and 18.75 Å, which are multiples of the $\pi-\pi$ stacking distance and are clearly originating from the lamellar aggregates. The peak at 5.25 Å does not appear among the nearest neighbors and is of unclear origin. A much larger proportion of rings are co-facial in the ES sample because their orientation is prevalently face-on. Still, all non-neighboring peaks are almost absent, indicating a less crystalline structure than the GS sample, in agreement with the observations from GIWAXS.

Finally, the surface roughness of the four samples was obtained from atomic force microscopy. The results, shown in Supplementary, Figure 12, indicate roughness of 20.0, 9.8, 2.7, and 4.2 nm for no, red, green, and blue light, respectively. This decrease in roughness points to an overall reduction of crystallinity and more homogeneous morphology as a consequence of light treatment, consistent with the GIWAXS and MD results.

\section*{Dynamics of Physical properties}
\label{Absorption Spectroscopy}
Normalized UV-VIS absorption spectra are shown in Figure \ref{fig:Absorbe}a with dashed lines indicating the absorption peaks $A_{1}$ and $A_{2}$. A broadening of the absorption into lower wavelength regime is observed for green and blue light treatment. It is thought that intrachain excitons in disordered chains cause such broadening \cite{JClark2009}. The absorption spectra are modelled to four Gaussian distributions, where the integrated intensities of $A_{1}$ and $A_{2}$ are shown in Figure \ref{fig:Absorbe}b, where the remaining two contributions are labeled as amorphous \cite{singH2017} (more details are found in Supplementary Figure 10). In Figure \ref{fig:Absorbe}b, the ratios of $A_{1}$ and $A_{2}$ are shown with respect to the no light treatment reference sample. Here, large and similar changes in absorption intensity, $A_{1}$ and $A_{2}$, are observed for green and blue light treatments. Unexpected, a slight change in the relative absorption in the red light treated film is observed, can be caused be the large bandwidth of the LED light. To interpret the observed changes in the absorption spectra, we adopt the weakly interacting H-aggregate model \cite{Spano2005ModelingFilms,JClark2009,p3htGold2016}, according to which the ratio between the fitted values of the absorption intensities $A_{1}$ and $A_{2}$ reads

\begin{equation}
\frac{A_{1}}{A_{2}} \approx \left( \frac{1-0.24W/E_p}{1+0.073W/E_p} \right)^2.    
\label{eq:AA}
\end{equation}
\begin{figure*}[ht]
\centering
  \includegraphics[width=\textwidth]{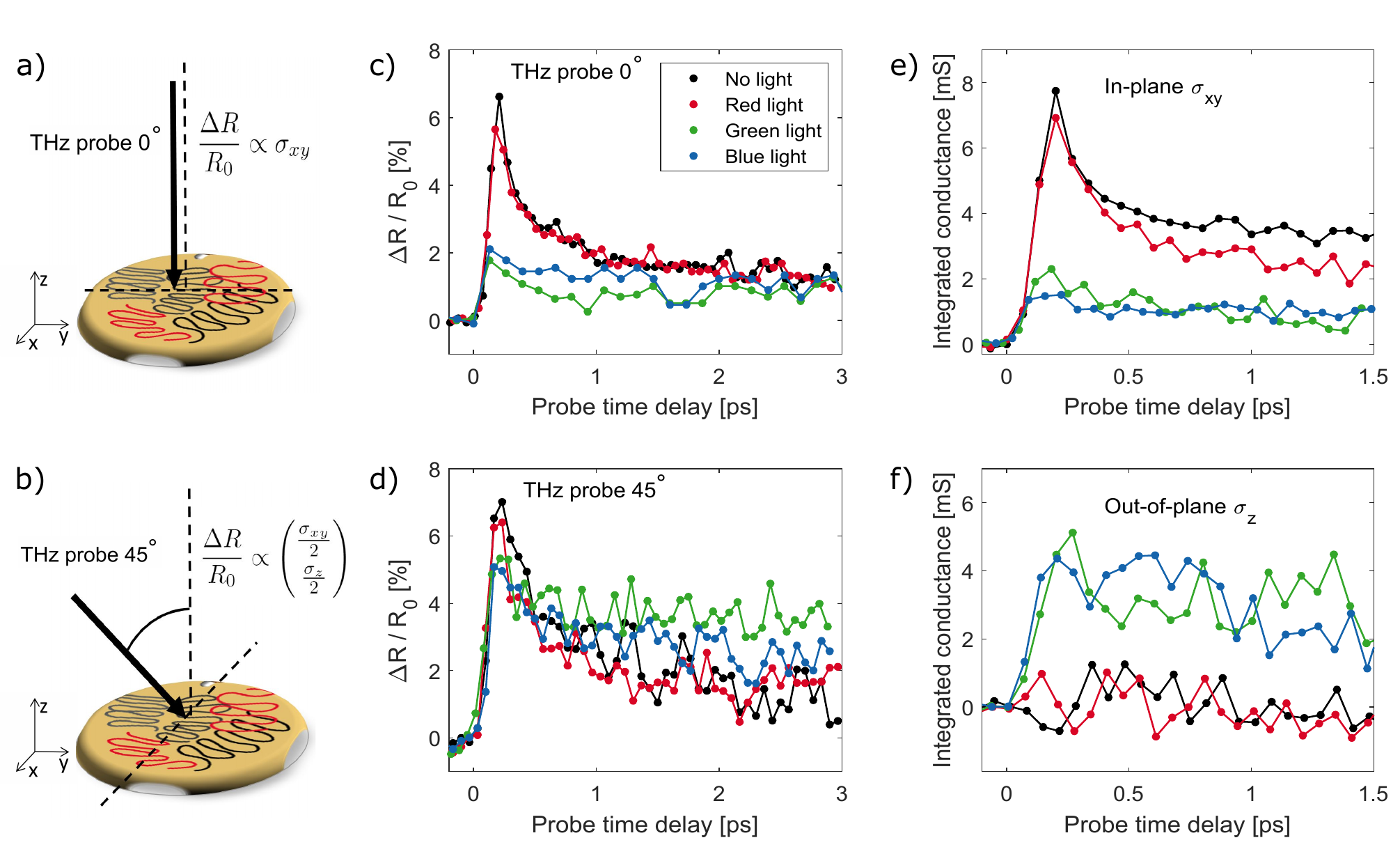}
  \caption{a), b) Illustration of the THz probe beam at 0$^{\circ}$ and 45$^{\circ}$ incidence angles respectively. At 0$^{\circ}$ the transient reflection $\Delta R / R_0$ probes the in-plane conductance $\sigma_{xy}$. At 45$^{\circ}$ $\Delta R / R_0$ probes a combination of in-plane and out-of-plane conductance $(\sigma_{xy}/2, \sigma_{z}/2)$. c) $\Delta R / R_0$ as a function of probe time delay for the P3HT thin films with different light treatments at normal incidence. d) $\Delta R / R_0$ as a function of probe time delay at 45° incidence. e) Spectrally integrated in-plane sheet conductivity extracted from a full 2D scan at normal incidence. f) Spectrally integrated out-of-plane sheet conductivity extracted from a full 2D scan at 45$^{\circ}$ incidence.}
  \label{fgr:THz}
\end{figure*}
In eq. (\ref{eq:AA}), $W=4J$ is the free-exciton bandwidth of the aggregates, where $J$ is the Coulombic inter-chain electronic coupling. $E_p$ = 0.18 eV is the energy of the C=C symmetric stretching, assumed to be the main intramolecular vibration coupled to the electronic transition \cite{Spano2005ModelingFilms,Spano2014} and the corresponding Huang-Ryhs factor is assumed to be 1. The ratios ${A_{1}}/{A_{2}}$ and the corresponding values of $W$ obtained from eq. (\ref{eq:AA}) are shown in Figure \ref{fig:Absorbe}c, and the exact numbers and errors are provided in Supplementary Table 2. The inter-chain coupling $J$, and consequently $W$, is known to decrease with increasing conjugation length\cite{Spano2014}. The increase of $W$, observed when going from no and red light to green and blue light treatment, indicates a decrease of conjugation length consistent with a more disordered morphology with smaller and fewer aggregates.


\subsection*{Correlation of structure and conductivity}

Time-resolved terahertz spectroscopy (TRTS) is a powerful tool for investigating ultrafast charge carrier transportation dynamics in organic materials at a characteristic length scale on the order of 10 nm \cite{Cooke2012,THz2012}. A THz wave is always polarized perpendicularly to its propagation direction, and therefore only charge carrier motion in directions perpendicular to the THz beam are probed. To probe both in-plane and out-of-plane charge carrier dynamics,  reflection-mode TRTS measurements were performed with two different incidence angles onto the P3HT films. At normal incidence angle only the in-plane photo-conductivity $\sigma_{xy}$ is picked up; at 45$^{\circ}$ incidence angle both in-plane and out-of-plane conductivities $\sigma_{z}$ can be extracted from the total conductivity $\sigma = \sigma_{z}\sin\theta + \sigma_{xy}\cos\theta$, as illustrated in Figure \ref{fgr:THz}a and \ref{fgr:THz}b.

In Figures \ref{fgr:THz}c and \ref{fgr:THz}d., the time evolutions of the transient THz wave reflection $\Delta R / R_0$, caused by photoexcitation of the film, were measured for all four samples at both 0$^{\circ}$ and 45$^{\circ}$. At normal incidence (probing in-plane mobility), the no light and red light treatment yield much stronger transient responses, with fast trapping of hot carriers. In contrast, the green and blue light treatment yield a much weaker in-plane response, and the trapping is more moderate. Probing with a 45° configuration, the green light and blue light treatments lead to a dramatic increase in transient response. In contrast, the no light and red light only show minor changes compared to the normal incidence configuration.

Full 2D TRTS scans (see Supplementary Figures 17-19) were performed to get explicit information about the photo-conductivity of the films. Sheet conductances were obtained by spectral integration of the 2D scans in the 2 to 12 THz frequency range. The in-plane conductance $\sigma_{xy}$, obtained from the normal incidence scan, is shown in Figure \ref{fgr:THz}e. The out-of-plane conductance component extracted from the 45$^{\circ}$ scan, $\sigma_{z}$, is shown in Figure \ref{fgr:THz}f. The contrast between the two groups of light treatment is pronounced: whereas the no and red light films predominantly exhibit in-plane photo-conductivity, the green and blue light treated films strongly favor out-of-plane charge transport. Since conductance in polymers is limited by inter-chain transport, i.e., along the $\pi$-stacking direction in semi-crystalline P3HT, this clear result strongly confirms the prevalence of aggregates oriented edge-on in the no and red light films and face-on in the green and blue light films, as observed by our GIWAXS measurements and MD simulations.

\section*{Technological Impact} 

The experimental and computational results presented in this paper demonstrate that exciting P3HT with visible light during deposition serves as a tool for manipulating the packing behavior of P3HT and can be used to modify the final thin film morphology. It can be considered a new processing parameter for achieving the desired performance of organic thin films. Particularly, the capability to increase the out-of-plane mobility by light treatment can be used for transistor applications where directional mobility and patterning is essential \cite{Edberg2016}. In a poor solvent for P3HT, such as ortho-xylene, the light treatment will not prevent P3HT from forming aggregates over time (see Supplementary, Figure 4). Investigating the many aspects of optimization to determine the optimal wavelength for treating a polymer in combination with temperature and solvents will be a full study of its own.
Furthermore, optimizing OPV devices comes with an additional challenge when two constituents require optimization processes for the right crystal packing and domain size. For the specific combination P3HT:O-IDTBR, the domain size decreases with light treatment and still results in a decrease in power conversion efficiency, as shown in Supplementary, Figures 13-15. However, understanding how light treatment during fabrication influences the final morphology of a film can enable major improvements for specific materials systems or other technologies in flexible electronics.

In conclusion, we report a method to manipulate the morphology of P3HT thin films through illumination with visible LED light during roll-to-roll slot-die coating. Optical polymer excitation temporarily constrains large sections of the chains into a planar geometry that is the minimum of the excited state potential energy surface. This structural effect is strong enough to affect the aggregation behavior and, thereby, the final morphology of the P3HT film. The light-treated films are less crystalline overall, display a higher degree of face-on orientation, shorter conjugation length, and a change of the unit cell dimensions with less efficient packing. Consequently, the in-plane photo-conductivity decreases and the out-of-plane conductivity increases drastically with light treatment.


\ack{The authors acknowledge financial support from the H2020 European Research Council through the SEEWHI Consolidator grant, ERC-2015-CoG-681881 and from the Independent Research Fund Denmark, grant no. 0200-00001B. Danscatt for travel expenses during beam times: DESY proposal number 15559230912. SAXS measurements were carried out at the P03/MiNaXS beamline, PETRA III at DESY. Neutron beam time at MLZ, proposal-15551". Some WAXS experiments were carried
out at the cSAXS beamline, Paul Scherrer Institute, Switzerland, proposal number 20182246.  Deuterated SD-P3HT was synthesized at the Center for Nanophase Materials Sciences, which is a DOE Office of Science User Facility (CNMS Proposal ID: CNMS2020-R-00546), Kristian Larsen for countless hours of technical support.}



\printbibliography

\newpage
\section{Materials and Methods}
\label{Materials and Methods}
All P3HT, Poly(3-hexylthiophene-2,5-diyl), in this work were purchased from Osilla: Lot number M1011, with a molecular weight Mw = 60.15 kDa, and a regioregularity of RR = 96.76 $\%$. All solutions were prepared with dichlorobenzene ($>$99.0 $\%$) with 20 mg/mL of P3HT and were stirred for 12 hours at 60 $^{\circ}$C until fully dissolved. A deuterated version, SD-P3HT, was synthesized at the Center for Nanophase Materials Sciences and the same batch as presented by \textit{Kunlun et. al.}\cite{kunlun2017}. O-IDTBR is purchased from 1-Material Inc. Silicon (100) and glass substrates were initially cleaned for 30 min in an ultrasonic bath: 10 min in isopropanol, 10 min in acetone, and 10 min in demineralized water. The slot-die coating procedure was done by mounting the rigid substrate of either silicon or glass on top of PET that were then running roll-to-roll while coating. This procedure can be found in a video article \cite{mksJove}, where the pumping rate is 0.08 mL/min, the speed of the substrate is 1 cm/min, and the width of the film 1 cm. We obtained a dry film of P3HT with thickness 210 nm ($\pm$ 20 nm) and length limited by the substrate dimensions, here 5-10 cm. All samples were coated at room temperature (approximately 22 $^{\circ}$C) in a fume hood. When the full length of the substrate had been coated, the syringe pump was stopped and the substrate kept moving the sample an additional 7 cm to the rest station. At the rest station the samples were exposed to either red, green, blue, or no light for 10 minutes. All other lights in the lab were turned off during the samples' preparation. For the light exposed samples three LEDs (3W LED, on PCB) with 1.5 cm spacing were placed 4 cm above the samples. The LEDs were  connected in series and a direct current of 1 Ampere and a voltage of 9 volts was applied. For all samples, the slot-die head was placed 7 cm from the sustrate. The temperature of the substrate remained constant during the light treatment. 

\subsection{Computational methodology} \label{CompMeth}
Classical molecular dynamics (MD) simulations were performed to generate plausible morphologies of solution-processed P3HT films deposited in absence and in presence of visible light illumination. To simulate the effect of light-induced electronic excitation on molecular geometry and therefore film morphology we derived two distinct force fields for the GS and ES of P3HT. A force field for the GS of P3HT was derived from the existing OPLS-AA (Optimized Potentials for Liquid Simulations - All Atom) force field and refined where needed with \textit{ab initio} density functional theory (DFT) calculations on 3MT (3-methyl-thiophene) oligomers. The atomic charges and a few structural parameters (inter-monomer bond length, angles within the backbone) were obtained from geometry optimization of a P3MT 8-mer at the B3LYP/6-311++G(d,p) level of theory using Gaussian 16\cite{g16}.
The atomic charges were obtained by fitting to the electrostatic potential (ESP) with the CHELPG scheme implemented in Gaussian 16 and then rescaled to yield net zero charge on internal monomers and equal charges of opposite sign on the terminal monomers. The proper dihedral angles between monomers were modelled with a Ryckaert-Bellemans (RB) functional form. The RB coefficients were obtained by fitting to the torsional profile of a 3MT dimer calculated at the $\omega$B97X-D/6-311++G(d,p) level of theory. The torsional profile (energy as a function of the dihedral angle) was calculated with a relaxed scan, i.e. a series of geometry optimizations with the dihedral angle constrained to fixed values between 0$^{\circ}$ and 180 $^{\circ}$.
To simulate the effect of illumination with classical MD, we built a different force field for the first ES of P3HT. The ESP atomic charges were obtained from geometry optimization of the first ES of a 3MT 8-mer with TDDFT/B3LYP/6-311++G(d,p) and then rescaled with the same procedure as for the GS. The inter-monomer dihedral angle parameters were obtained with the same procedure as for the GS, but the torsional profile was obtained from a relaxed scan performed on the first ES of a 3MT dimer calculated at the TDDFT/$\omega$B97X-D/6-311++G(d,p) level of theory. All other ES force field parameters are identical to the GS force field.
Coarse-grained (CG) force fields, where several atoms are represented by one larger particle (bead), have the goal of reducing the number of particles in the simulation and thereby dramatically increase the system size and/or simulation length achievable with given computational resources. Here we adopt a MARTINI 3.0 \cite{Souza2021} CG model of P3HT \cite{AndersPHD} \cite{asg2021} based on the OPLS-AA force field described above. In the CG model the thiophene ring is represented by three tiny beads of type TC6, TC5 and TC5 with constraints between them to keep the ring rigid. One virtual site (VS) is placed at the center of geometry (COG) of the ring. The alkyl chain is connected to the ring by a harmonic bond and is represented by two small beads of type SC2, also connected by a harmonic bond. The chlorobenzene molecule is represented by three beads with constrained bonds: two of type TC5 and one of type SX3. A visual representation of the CG mapping is shown in Supplementary Figure 1. The GS and ES CG force fields differ only by the inter-monomer dihedral angle parameters, which were taken from the OPLS-AA force fields described above. The CG model and force field of chlorobenzene was described in ref. \cite{AndersPHD}.

P3HT thin film morphologies were generated using a previously published solvent evaporation scheme based on CG MD simulations where the solvent molecules are progressively removed from the simulation box until none are left, mimicking the drying of a thin film deposited in a slot-die coating process \cite{Lee2014, Riccardo2017, Gertsen2020}. The same simulation was run both with the GS and the ES CG force fields, with all other conditions identical. The initial configuration of the system was prepared by randomly placing 420 P3HT 48-mers in a periodic simulation box of initial dimensions 25 x 25 x 136.7 nm and solvating them with 469725 chlorobenzene molecules, yielding a concentration of around 40 mg/mL. The system was first equilibrated for 0.5 ns in an NVT ensemble and then for 4.0 ns in an NPT ensemble. After this, the evaporation run was started: at each step, 1.25 $\%$ of the remaining solvent molecules, randomly selected, were removed from the box. After reaching the near-linear regime, i.e. when less than 0.0125 $\%$of the initial amount of solvent was removed at each step, the evaporation was continued linearly (removing the same number of molecules each step) until a dry film (without solvent) was obtained, amounting to a total of 264 steps. At each step, after removing the solvent the system was equilibrated for 0.5 ns in an NVT ensemble (Berendsen thermostat, coupling constant $\tau$ = 2 ps) and for 4.0 ns in an NPT ensemble (Berendsen thermostat, $\tau$ = 2 ps, and Berendsen barostat, $\tau$ = 4 ps) before undergoing a production run of 3.0 ns in an NPT ensemble (Berendsen thermostat, $\tau$ = 2 ps, Parrinello-Rahman barostat, $\tau$ = 15 ps), adding up to a total evaporation time of around 2 $\mu$s. A 20 fs time step was used for the leap-frog integrator, and semi-isotropic pressure coupling was applied with a pressure of 1 bar and a compressibility of \num{4.5e-5} bar$^{-1}$ in the \textit{z}-direction and of 0.0 bar$^{-1}$ in the textit{x}- and \textit{y}-directions to facilitate shrinking of the box only in the \textit{z}-direction, hence mimicking the conditions in a drying thin film. No charges were present in the systems, meaning that the electrostatics could be ignored. The van der Waals interactions were treated with the potential-shift Verlet scheme with a cut-off of 1.1 nm as recommended for MARTINI force fields.
From each of the two final thin film morphologies, two aggregates of 8 48-mers were selected for the calculation of UV-Vis absorption energies. First, the aggregates were back-mapped to atomistic resolution with the \textit{backward} method described in ref \cite{Wassenaar2014} using an initial random displacement ("kick") of 0.35 nm. The geometries of the resulting back-mapped aggregates were then optimized with GROMACS using the GS OPLS-AA force field and non-aggregated portions of the molecules were removed to obtain 8 24-mers (6 24-mers and 1 48-mer in one of the ES aggregates). To further reduce the atom count, the hexyl side chains were substituted with methyls, since \textit{sp}3 side chains have little influence on the electronic structure of the excited states. Excited state calculations on systems of this size ($\approx$1900 atoms) are not feasible with TDDFT, therefore we computed the single point electronic structure with the semi-empirical DFTB3 method with the 3ob-3-1 parameter set \cite{Gaus2013,Gaus2014} using the DFTB engine of the AMS 2020 program suite\cite{r_ruger_ams_2020,Velde2001}. The energies and oscillator strengths of the lowest 60 singlet excitations were calculated with the TD-DFTB method \cite{Ruger2015} as implemented in AMS DFTB. The computed absorption spectra shown in Supplementary, Figure 11, were obtained by combining the excitations of two aggregates from each of the thin film morphologies and convoluted with Gaussian functions of width 3.0 nm and weights corresponding to the oscillator strengths.

\subsection{Experimental methodology}
\subsubsection{Absorption and External quantum efficiency}
Absorption spectroscopy and External quantum efficiencies (EQE) as function of incident wavelength where obtained by using a QEX10 system (PV Measurements Inc.). For all measurements the incident wavelength were probed from 300 to 900 nm in steps of 5 nm. The system is calibrated with a Si photo diode.

\subsubsection{GIWAXS}
To perform Grazing Incidence Wide-Angle X-ray Scattering (GIWAXS) measurements, a laboratory setup was used (Xeuss 3.0 from Xenocs S.A.). Here, supplied with a microfocus copper source, Cu K $_{\alpha}$ radiation (waveleng ($\lambda$) = 1.5418 Å) is focused and monochromatized with a two-dimensional single reflection multilayer optic and collimated with scatterless slits. The silicon substrate surface was aligned at a grazing incident angle of 0.18° with respect to the incoming X-ray beam. The scattered X-ray were detected on an Eiger 4M single photon counting detector, with 75 µm pixels  (DECTRIS), 90.0 mm from the sample. Conversion of the raw 2D data to reciprocal space coordinates was performed with the SimDiffraction MATLAB script package \cite{DagBrei2008}.

\subsubsection{Time-resolved THz spectroscopy}
For the transient optical-pump THz-probe spectroscopy investigation, 2 mJ, 40 fs pulses with 1 kHz repetition rate, 800-nm central wavelength from a Ti:sapphire laser amplifier (Spectra-Physics, Spitfire Ace) are split into three beams for terahertz wave generation, waveform detection, and for generating the optical excitation pulses at 400 nm, respectively. The ultra-broadband THz pulse generation is realized by 2-colour fs laser induced air-plasma process in the focal zone of the fundamental driving laser beam (800 nm) and its second harmonic beam. To preserve the ultra-broad THz bandwidth from the generation process, an air-biased coherent detection (ABCD) scheme was implemented. The 400-nm optical excitation beam was projected onto the sample with 1.4 mm diameter (full width half maximum) spot size in the normal incident case. For the 45° incidence configuration, the spot onto the tilted sample is larger (1.414 times). We use same pump fluence of 585 $\mu J/cm^2$ for all the measurements, which is slightly below the damage threshold of the polymer films.  
\subsubsection{Atomic Force Microscopy}
To obtain the surface morphologies of the layers, the scanning probe microscopy (DualScopeTM scanner DS 95-50 with a DualScopeTM controller C-26, DME) was used with Si tips (ArrowTM NCR, NanoWorld) at room temperature. The spring constant of the tip is ~42 N/m, and the curvature radius is below 10 nm. Thin film where prepared on Si (100) wafers and cut in appropriate sizes ($1 \times 1$  cm$^2$). 

\subsubsection{SAXS}
Small Angle X-ray Scattering (SAXS) in transmission mode measurements where performed at DESY, P03 at the microfocused end-station \cite{P03Micro}. The energy was chosen to be 11.2 keV and a sample to detector distance calibrated with AgBeh to be at 3900 mm. The scattered X-ray are detected by a Pilatus 300k pixel detector. To load the data and perform flat-field correction, DPDAK software was used \cite{DPDAK}.

\subsubsection{GISANS}
Grazing Incidence Small-Angle Neutron Scattering (GISANS) measurements where performed at MARIA beam line at FRM2 \cite{MariaMLZ2018}, with a probing wavelength of 10 Å unpolarized neutrons with a sample to detector distance of 1910 mm. The incident angle where optimized at the reflected neutron beam at 0.52$^{\circ}$.

\subsubsection{Organic Solar Cells}
The fabrication procedure is outside the scope of the paper, and can be found in \cite{marcial2020}. The Device structure is the PET/ITO/ZnO/P3HT:O-IDTBR/MoO/Ag. While coating the active layer (P3HT:O-IDTBR), the inks where exposed to either no, red, green, and blue light. All current density-voltage (J -V) curves are measured using a solar simulator with a xenon lamp as the light source, which is calibrated by a certified reference cell (monocrystalline silicon certified by Fraunhofer) under AM1.5G illumination, 1000 W/m2 intensity, and acquired by a Keithley 2400 source meter (25 mV step reverse to forward). No current is obtained outside the solar-cell area, as the flextrode is cut to ensure 1 cm$^2$ devices.

\end{document}